# Nanoantenna properties of graphene membrane. Quantum theory.

## N.E. Firsova[1,2] and Yu.A. Firsov[3]

[1]Institute for Problems of Mechanical Engineering, the Russian Academy of Sciences, St. Petersburg 199178, Bolsoy Prospect V.O. 61,Russia,

[2]State Polytechnical University, St. Petersburg 195231, Politechnicheskaya 29, Russia,

**email:** nef2@mail.ru

[3]A.F.Ioffe Physical-Technical Institute, the Russian Academy of Sciences, St.Petersburg, 194021, Russia,

**email:** yuafirsov@rambler.ru

**Abstract.**

The graphene membrane irradiated by weak activating alternative electric field in terahertz range was considered. The quantum approach based on the time-dependent density matrix method was used. The exact solution was obtained for graphene membrane density matrix equation in linear on the external field approximation. The graphene electromagnetic response was studied i.e. the formulae for the induced current and conductivity were found and analyzed neglecting loss. The obtained formula for the conductivity showed that the graphene membrane was an oscillating contour and its fundamental eigen frequency coincided with a singularity point of the conductivity. This formula allowed us to calculate the graphene membrane inductivity and the capacitance. So the graphene membrane could be used as an antenna or a transistor. It was shown that its eigen frequency could be tuned by doping as its value was found to depend on electrons concentration. It was obtained that the eigen frequency could be tuned in a rather large frequency range 1-100 terahertz as electrons concentration in graphene may differ considerably. The found dependence on concentration correlates with experiments. It would be useful to take the obtained results into account when constructing devices containing graphene membrane nanoantenna allowing wireless communications among nanosystems. This could be promising research area of energy harvesting applications.

**1.Introduction.**

After the graphene discovery [1-2] there appeared many interesting experimental results (see for instance [3]) which were explained rather well in the framework of 2D (flatland) model. However after a while it was found that the monolayer graphene surface was strongly corrugated [4]. This surface curvature under external time depending field generates "breathing" metrics which results in certain new effects. These effects proved to be essential and were for the first time analyzed in [5-7]. It was shown in [5] that the quality factor of the graphene nanoresonator essentially depends on graphene membrane metrics generating new loss mechanism based on synthetic electric fields. These fields arise due to the "breathing" graphene membrane metrics activated by the external periodic electric field. In [6-7] it was found that graphene electromagnetic response i.e. induced currents were curved, the curvature essentially depending on the graphene membrane surface metrics. Results of [5-7] explain experimental data. The methods used there were based on the modified Boltzmann equation and on taking into consideration so called synthetic electric fields. Nowadays investigation of   feature-rich electronic properties in rippled graphene become more and more significant, [8], especially in connection with the appearance of new 2D-materials (silicone, germanene, stanene, phosphorene and so on).

Our study in [5-7] was done in the framework of quasiclassics. Now our goal is to study graphene surface corrugation influence in nonstationary processes using quantum mechanics approach which seem to be more adequate in the case of graphene membrane. In this paper we use nonstationary density matrix method. And this quantum approach to the nonstationary physics in graphene allows us to find out new quantum effects and to obtain more adequate picture of graphene than the strain model, using elastic theory which was explained in detail in [9]. As we take flatland model as an unperturbed Hamiltonian the contribution of corrugations will be considered as $H_{int}$ (the same way as external nonstationary field). So first of all we should consider the problem in the framework of flatland model. As a first step we also neglect e-e-interactions and any kind of loss. It proved to be that the consideration even of this simplified unperturbed problem gives essential corrections to the physical picture compared to earlier papers (see [10-11], more in detail will be explained below). In this paper we obtain induced current and graphene membrane conductivity as a function of the alternative external electric field frequency. The found formula for conductivity allows us to determine the fundamental eigen frequency of the graphene membrane, its eigen inductivity and capacity. So we show that graphene membrane can be used as an antenna radiating a frequency which can be tuned in certain range. The obtained results correlate with experiments. We think that the found formulae can be used in different kinds of applications, especially for wireless communications among nanosystems. The corrections connected with the inevitable surface corrugations will be described in our next paper.

**2. Statement of the problem.** We consider the current induced in graphene membrane irradiated by the weak external complex activating electric field $\vec{E}_0 e^{i\omega t}$. We assume this field to be homogeneous which is reasonable as we consider terahertz frequency range (in terahertz frequency range the wavelength is usually much more than graphene membrane size). We solve the problem using the quantum approach and density matrix. The Hamiltonian has a form

$$\widehat{H} = \widehat{H}_0 + \widehat{H}_{int}, \qquad (1)$$

where

$$\widehat{H}_0(\vec{k}) = \hbar v_F \vec{k} \cdot \vec{\sigma} = \hbar v_F (k_1 \sigma_1 + k_2 \sigma_2) = \hbar v_F k \begin{pmatrix} 0 & e^{-i\varphi} \\ e^{i\varphi} & 0 \end{pmatrix}, \qquad (2)$$

$$\sigma_1 = \begin{pmatrix} 0 & 1 \\ 1 & 0 \end{pmatrix}, \quad \sigma_2 = \begin{pmatrix} 0 & -i \\ i & 0 \end{pmatrix}, \qquad \vec{k} = k_1 + ik_2 = ke^{i\varphi}$$

and

$$\widehat{H}_{int}(\vec{k}) = e \frac{v_F}{c} (A_1 \sigma_1 + A_2 \sigma_2) \qquad (3)$$

Here $\vec{A} = (A_1, A_2)$ is a vector potential related to the external electric field $\vec{E}_0 e^{i\omega t}$ that is

$$\frac{1}{c} \frac{\partial \vec{A}}{\partial t} = \vec{E}_0 e^{i\omega t}, \quad \vec{A} = c \frac{e^{i\omega t}}{i\omega} \vec{E}_0, \quad A_{1,2} = c \frac{e^{i\omega t}}{i\omega} E_{01,02}. \qquad (4)$$

$$E_{01} = E_0 \cos\theta, \qquad E_{02} = E_0 \sin\theta$$

Substituting vector potential coordinates (4) into (3) we get

$$\widehat{H}_{int}(\vec{k}) = \frac{ev_F}{i\omega} e^{i\omega t}[E_{01}\sigma_1 + E_{02}\sigma_2] = \frac{ev_F}{i\omega} E_0 e^{i\omega t} \begin{pmatrix} 0 & e^{-i\theta} \\ e^{i\theta} & 0 \end{pmatrix} \qquad (5)$$

We shall calculate the induced current using the density matrix $\hat{\rho}$ which is a solution of the equation

$$\frac{d\hat{\rho}}{dt} = -\frac{i}{\hbar}[\widehat{H},\hat{\rho}]. \qquad (6)$$

Then the coordinates of the induced current operator $\vec{j} = (\hat{j}_1, \hat{j}_2)$, $\hat{j}_{1,2} = ev_F \sigma_{1,2}$, will be equal to

$$j_{x,y} = ev_F Sp(\hat{\rho}\sigma_{x,y}) \qquad (7)$$

And the total current will be

$$j_{1,2}^{tot} = \int_0^\infty k\,dk \int_0^{2\pi} j_{1,2}\,d\varphi \qquad (8)$$

We shall obtain the exact solution for the equation (6) in linear approximation which is reasonable as we consider weak external electric field. Using the obtained formula we find the induced current from (7). Hence we get the formula for conductivity which is imaginary in the approximation we consider (in our present calculations we neglect e-e interactions and any kind of loss). We think that the obtained formulae after generalization with taking into account the loss, e-e interactions and inevitably existing corrugations will be useful in applications. The taking into consideration the obtained results when constructing devices containing graphene membranes would allow to increase the exactitude of measurements.

### 3. Exact solution of the density matrix equation in linear approximation

We consider the density matrix equation (6) in linear approximation for the case of the activating field $\vec{E}_0 e^{i\omega t}$. Such approximation is reasonable as we consider weak activating electric field. The linear approximation of the equation (6) will be as follows

$$\frac{d\hat{\rho}_{int}}{dt} = -\frac{i}{\hbar}\left([\widehat{H}_0, \hat{\rho}_{int}] + [\widehat{H}_{int}, \hat{\rho}_0]\right) \qquad (9)$$

Here $\hat{\rho}_0$ is a Fermi-Dirac distribution corresponding to the unperturbed case and $\hat{\rho}_{int} = \hat{\rho} - \hat{\rho}_0$ is an unknown matrix. To find this unknown matrix we shall transform Hamiltonian $\widehat{H}_0$ to the diagonal form. It is not difficult to check that

$$\widehat{H}_0(\vec{k}) = \hbar v_F k U^{-1} \sigma_3 U \qquad (10)$$

where

$$\sigma_3 = \begin{pmatrix} 1 & 0 \\ 0 & -1 \end{pmatrix}, \quad U = \frac{1}{\sqrt{2}}\begin{pmatrix} e^{i\varphi} & 1 \\ 1 & -e^{-i\varphi} \end{pmatrix}, \quad U^{-1} = \frac{1}{\sqrt{2}}\begin{pmatrix} e^{-i\varphi} & 1 \\ 1 & -e^{i\varphi} \end{pmatrix} \qquad (11)$$

Then we have

$$\hat{\rho}_0 = U^{-1}\hat{f}\,U = U^{-1}\begin{pmatrix} f_{FD}(\varepsilon) & 0 \\ 0 & f_{FD}(-\varepsilon) \end{pmatrix}U, \qquad (12)$$

$$f_{FD}(\pm\varepsilon) = [1 + exp\{\beta(\pm\varepsilon - \mu)\}]^{-1} = [1 + exp\{\beta(\pm\hbar v_F k - \mu)\}]^{-1}, \; \beta = \frac{1}{k_B T}. \qquad (13)$$

Now we solve the equation (9) where $\hat{\rho}_{int}$ is unknown. We find the solution

$$\hat{\rho}_{int} = \frac{1}{2}\frac{ev_F}{\hbar\omega}\sin(\theta - \varphi)\,\Delta\begin{pmatrix} \gamma_{11} & \gamma_{12} \\ \gamma_{21} & \gamma_{22} \end{pmatrix}E_0 e^{i\omega t}, \qquad (14)$$

$$\Delta = f_{FD}(\varepsilon) - f_{FD}(-\varepsilon),$$

where

$$\gamma_{11} = [(2v_F k + \omega)^{-1} - (2v_F k - \omega)^{-1}], \qquad (15)$$

$$\gamma_{12} = -e^{-i\varphi}[(2v_F k + \omega)^{-1} + (2v_F k - \omega)^{-1}], \qquad (16)$$

$$\gamma_{21} = e^{i\varphi}[(2v_F k + \omega)^{-1} + (2v_F k - \omega)^{-1}], \qquad (17)$$

$$\gamma_{22} = -[(2v_F k + \omega)^{-1} - (2v_F k - \omega)^{-1}]. \qquad (18)$$

**4. Induced current, conductivity. Eigen inductivity, capacitance and frequency for the graphene membrane.**

We have found the exact solution of the density matrix equation in linear approximation (9) for the external electric field $\vec{E}_0 e^{i\omega t}$. So we have

$$\hat{\rho} = \hat{\rho}_0 + \hat{\rho}_{int}, \qquad (19)$$

where $\hat{\rho}_{int}$ we see in (14)-(18). Now we can calculate the induced current using the formula (7). So we have for the first coordinate of the current (in momentum representation)

$$j_1 = ev_F Sp(\hat{\rho}_{int}\sigma_1), \qquad (20)$$

since the first term $\hat{\rho}_0$ (as one can easily check) does not generate any current. Substituting (14)-(18) into (20) we obtain the first coordinate of the current

$$j_1 = \frac{e^2 v_F^2}{2\hbar\omega}\sin(\theta - \varphi)\,\Delta\,Sp\left(\begin{pmatrix} \gamma_{12} & \gamma_{11} \\ \gamma_{22} & \gamma_{21} \end{pmatrix}\right)E_0 e^{i\omega t}$$

Using (16)-(17) we find

$$j_1 = i\frac{e^2 v_F^2}{\hbar\omega}\sin(\theta - \varphi)\sin\varphi\,\Delta[(2v_F k + \omega)^{-1} + (2v_F k - \omega)^{-1}]\,E_0 e^{i\omega t} \qquad (21)$$

Now we can find the total value of the first coordinate of the current from the formula

$$j_1^{tot} = \int_0^\infty k\,dk \int_0^{2\pi} j_1\,d\varphi \qquad (22)$$

Since

$$\int_0^{2\pi} \sin(\theta - \varphi) \sin \varphi \, d\varphi = -\pi \cos \theta$$

we obtain

$$j_1^{tot} = -i\pi \frac{e^2 v_F}{2\hbar\omega} E_{01} e^{i\omega t} \int_0^\infty \Delta \left[\left(k + \frac{\omega}{2v_F}\right)^{-1} + \left(k - \frac{\omega}{2v_F}\right)^{-1}\right] k \, dk \qquad (23)$$

Integrating for low temperatures when $\Delta = 1$ inside $[0, k_F]$ and $\Delta = 0$ outside we obtain

$$j_1^{tot} = -2i\sigma_0 \left[\frac{E_F}{\hbar\omega} + \frac{1}{4} \ln \left|\frac{2E_F - \hbar\omega}{2E_F + \hbar\omega}\right|\right] E_{01} e^{i\omega t}, \quad \sigma_0 = \frac{\pi}{2} \frac{e^2}{\hbar}. \qquad (24)$$

The same way we can get the formula for the second coordinate. So we get

$$\sigma(\omega) = -i\pi \frac{e^2 v_F}{2\hbar\omega} \int_0^\infty \Delta \left[\left(k + \frac{\omega}{2v_F}\right)^{-1} + \left(k - \frac{\omega}{2v_F}\right)^{-1}\right] k \, dk \qquad (25)$$

and for $T = 0$ we find

$$\sigma_{FF}(\omega) = -2i\sigma_0 \left[\frac{E_F}{\hbar\omega} + \frac{1}{4} \ln \left|\frac{2E_F - \hbar\omega}{2E_F + \hbar\omega}\right|\right] \qquad (26)$$

Note that the formula similar to (26) for the case neglecting loss and e-e interactions was obtained in [10], Eq.(5.13)

$$\sigma_{GSC}(\omega) = \frac{2}{\pi} 2i\sigma_0 \left[\frac{|\mu|}{\hbar\omega} + \frac{1}{4} \ln \frac{2|\mu| - \hbar\omega}{2|\mu| + \hbar\omega}\right] \qquad (27)$$

We shall compare the results (26) and (27). The opposite sign in (27) is essential from physical reasons which will be explained below. Let us transform (26)

$$\sigma_{FF}(\omega) = -2i\sigma_0 \left[\frac{k_F v_F}{\omega} + \frac{1}{4} \ln \left|\frac{1 - \frac{\omega}{2k_F v_F}}{1 + \frac{\omega}{2k_F v_F}}\right|\right] \qquad (28)$$

For terahertz range $\frac{\omega}{2k_F v_F} \ll 1$ and consequently we have

$$\sigma_{FF}(\omega) \approx \frac{1}{i\omega L_{FF}} + i\omega C_{FF} \qquad (29)$$

where

$$L_{FF} = \frac{1}{2k_F v_F \sigma_0} = \frac{\hbar}{\pi e^2 k_F v_F} \qquad (30)$$

$$C_{FF} = \frac{\sigma_0}{2k_F v_F} = \frac{\pi e^2}{4\hbar k_F v_F} = \frac{\pi e^2}{4 E_F} \qquad (31)$$

Note that from the obtained formula (31) we find the capacitance of the mean graphene membrane area occupied by one electron

$$c_{FF} = \frac{\sqrt{\pi n}}{4} \frac{e^2}{\hbar v_F} \qquad (32)$$

From (29) we see that graphene membrane is a nanoantenna where inductivity $L_{FF}$ and capacitance $C_{FF}$ are joined in a parallel way. As to formula (27) it describes the oscillating contour with $L_{GSC} < 0$ and $C_{GSC} < 0$. So the formula (27) does not give an adequate physical picture.

Remark that in [12] the graphene membrane quantum capacitance was first measured. There was also suggested the formula for quantum capacitance based on a two-dimensional free electron gas model. The expression obtained in [12] coincides with (32) multiplied by $8/\pi$.

Note also that in [13] the suggestion was made that electrons in graphene must exhibit a nonzero mass when collectively excited. Using this notion the inertial acceleration of the electron collective mass and phase delay of the resulting current were considered. On the basis of this model the so-called kinetic inductance representing the reluctance of the collective mass to accelerate was introduced, calculated and measured. The obtained expression coincides with the formula (30) multiplied by $\pi^2$. Analyzing the formula for inductivity we see that $L_{FF} \sim n^{-1/2}$ that means that less is the concentration the more is the inductivity.

From (30), (31) we see that graphene membrane eigen frequency is as follows

$$\omega_{FF} = \frac{1}{\sqrt{L_{FF}C_{FF}}} = \frac{2E_F}{\hbar} = 2k_F v_F \tag{33}$$

We see also from the formula (26) that

$$\lim_{\omega \to \omega_{FF}} \sigma(\omega) = \infty \tag{34}$$

which means that graphene membrane is an oscillation contour with eigen frequency $\omega_{FF}$. Note that the formulae for inductivity and capacity from [12-13] as a matter of fact were obtained on the basis of different models. So if we used these formulae from [12-13] to find eigen frequency we would have obtained the value where the conductivity (26) does not have singularity. It means that the results obtained in [12-13] does not give possibility to find correctly graphene membrane eigen frequency.

**5. Discussion and conclusion**

We considered the graphene membrane electromagnetic response for weak periodic electric field in terahertz range in the framework of flatland model. We used quantum approach based on the nonstationary density matrix neglecting e-e interactions and any loss. We found exact solution of the equation for density matrix in linear approximation. On this basis we found the induced current and conductivity. For low temperatures we obtained simple formula for the conductivity. This formula gives purely imaginary value $\sigma_{FF}(\omega)$ which shows that graphene membrane has eigen inductivity, eigen capacitance and eigen frequency. The value of the found eigen frequency corresponds to the singularity of the conductivity $\sigma_{FF}(\omega)$. It means that graphene membrane is an antenna radiating this frequency or this is a resonant frequency of the graphene transister. However as our decision was obtained without taking into account e-e-interaction we are planning to find corrected value $\omega_{FF}^{int}$ of the eigen frequency. From our formula we see that

$$\omega_{FF} = 2v_F \sqrt{\pi n} \tag{35}$$

As far as concentration $n$ can vary from $10^9$ to $10^{13}$ we see that using doping by constant electric field (see for instance [14]) we can vary eigen frequency approximately from 10 to 100 terahertz. However the experiment [15] shows that graphene antenna radiates the frequency about 1-3 terahertz. It means that the corrected theory where we take into account e-e interactions should give $\omega_{FF}^{int}$ in this range. Such a big difference between $\omega_{FF}$ and $\omega_{FF}^{int}$ is expectable as in 2D e-e-interactions play essential role because in 2D we have $\alpha_{2D} = \frac{e^2}{\hbar v_F} \approx 2,2$ while in 3D $\alpha_{3D} = 1/137$.

Note that in [16] it was shown that the quantum capacitance decreases in presence of e-e interactions by 10-15% due to effective reduction of the density of states. However, generally it changes the result only quantitatively, retaining the same form $C_Q \sim \sqrt{n}$ as in the noninteracting model. That is why experimentally measured $C_Q$ can be successfully described in the noninteracting model but with higher value of velocity $v_F \approx 1.1 \times 10^8 sm/sec$ (see[16])

Note also that we can rewrite Eq. ([17],Eq(8)) as follows

$$\frac{\omega_p(q)}{\omega_{FF}} = \sqrt{2\alpha_{2D}} \sqrt{\frac{q}{k_F}} \qquad (36)$$

where $\omega_p(q)$ is Plasmon frequency and $q$ is Plasmon wave number. There are assumptions that $q \ll k_F$ and consequently $\omega_p \ll \omega_{FF}$. So we may assume that as a matter of fact the frequency $\omega_{FF}^{int}$ introduced above and the plasmonic frequency $\omega_p$ are the same thing. We are planning to analyze this question in the next paper.

To find correct value of the eigen frequency we should also take into account the loss and the graphene surface corrugations influence. In [5-7] where we studied induced current trajectories we found that they were curved and the curvature were determined by the graphene surface form. This result correlated with experiment where current patches were obtained and they proved to be curved [18] which can be explained only by our results. We also found there that the induced currents have phase delay depending on the point $(x, y)$. So it may happen that when we develop our quantum approach taking into account surface corrugations influence we shall find that the graphene membrane eigen inductivity, capacitance and frequency will prove to depend on the point $(x, y)$. This can lead to extra loss and lower quality factor as we demonstrated it considering graphene nanoresonator in [5]. We hope to answer these questions in our next papers which would be useful for constructing the improved quantum theory of graphene nanoantenna and its applications to wireless communications among nanosystems.